\ifdefined\reviewversion
  \PassOptionsToClass{preprint,linenumbers}{revtex4-2}
\else
  \PassOptionsToClass{reprint}{revtex4-2}
\fi
\documentclass[
  aps,
  prx,
  superscriptaddress,
  amsmath,
  amssymb,
  floatfix,
  longbibliography,
]{revtex4-2}

\usepackage{graphicx}
\graphicspath{{figures/}}
\usepackage{dcolumn}
\usepackage{bm}
\usepackage{microtype}
\usepackage{siunitx}
\usepackage[colorlinks=true,linkcolor=blue,citecolor=blue,urlcolor=blue]{hyperref}
\DeclareSIUnit{\belmilliwatt}{Bm}
\DeclareSIUnit{\dBm}{\deci\belmilliwatt}
\begin{document}

\title{Near-Quantum-Limited Readout for the Princeton Axion Search using a sub-GHz Kinetic Inductance Traveling Wave Parametric Amplifier}

\author{Joe Wiedemann}
\email{jw9239@princeton.edu}
\author{Roman Kolevatov}
\author{Jianyang Fu}
\author{Jo\"elle-Marie B\'egin}
\author{Nate Otto}
\author{Felix St-Amour}
\author{Lyman Page}
\affiliation{Department of Physics, Princeton University}
\author{Henry LeDuc}
\author{Peter Day}
\affiliation{Jet Propulsion Laboratory, California Institute of Technology}
\author{Saptarshi Chaudhuri}
\email{sc25@princeton.edu}
\affiliation{Department of Physics, Princeton University}

\date{\today}

\begin{abstract}
    Parametric amplifiers are a workhorse for near-quantum-limited measurements above $\SI{1}{\giga\hertz}$. However, amplifier development has been more limited below $\SI{1}{\giga\hertz}$. We present a kinetic inductance traveling-wave parametric amplifier (KI-TWPA) operating over 300-700 MHz, with gain exceeding $\SI{30}{\deci\bel}$ and pump powers of a few hundred nanowatts. The system noise performance, as measured by a variable-temperature noise source, is 2.5-3.5 quanta. This device is a versatile new tool for precision measurements of sub-GHz signals, with applications including readout of semiconductor spin qubits, superconducting detector arrays, and searches for new physics beyond the Standard Model. We characterize aspects of the amplifier performance relevant to axion searches, including gain drift, compression point, amplifier noise at various operating temperatures, readout with an input circulator, and magnetic field insensitivity.
\end{abstract}


\maketitle

\section{Introduction}
\label{sec:introduction}

Parametric amplifiers are an essential component for quantum information science, astrophysics, cosmology, and dark matter searches \cite{krantz2019quantum, zobrist2019wide, malnou2023improved, ranzani2018kinetic, ramanathan2023wideband}. Several parametric amplifier architectures operating above $\SI{1}{\giga\hertz}$ have demonstrated performance approaching the Standard Quantum Limit (SQL) \cite{aumentado2020superconducting}. 
Further, Josephson Parametric Amplifiers (JPAs), Josephson Parametric Converters (JPCs), and Josephson Traveling-Wave Parametric Amplifiers (JTWPAs) have been used to generate single- and two-mode-squeezed states for measurements evading the SQL \cite{castellanos2007widely,castellanos2008amplification,bergeal2010phase,backes2021quantum,jewell2023new,qiu2023broadband}. Outside of Josephson junctions, another means to generate the non-linearity required for parametric amplification is kinetic inductance. Kinetic Inductance Traveling Wave Parametric Amplifiers (KI-TWPAs) operating at several GHz have recently demonstrated noise approaching that of the best JTWPAs and offer similar, octave-scale bandwidth \cite{macklin2015near,white2015traveling,wang2025high,Eom2012,faramarzi20244, howe2026kinetic}. KI-TWPAs come with the potential advantage of higher dynamic range and greater immunity to magnetic fields \cite{janssen2026temperature,boutan2026operation}.

Over the $\sim \SI{200}{\mega \hertz}-\SI{1}{\giga\hertz}$ frequency range, the development of quantum-limited microwave amplifiers has been more limited \footnote{Below $\sim \SI{200}{\mega \hertz}$, dc SQUIDs provide a mature option for near-quantum-limited amplification. See Ref. \cite{ankel2025noise} for a detailed discussion.}. Nonetheless, a large number of applications would benefit from further development. High-fidelity, single-shot spin-qubit readout has been accomplished with JPAs operating over $\qtyrange{500}{800}{\mega\hertz}$ \cite{schaal2020fast, oakes2023fast}. However, owing to their resonant nature, the JPAs are limited in their instantaneous bandwidth and saturation power, e.g. \citet{oakes2023fast}: a $\SI{3}{\deci\bel}$ bandwidth of $\approx \SI{8}{\mega\hertz}$ and an input $\SI{1}{\deci\bel}$ compression point of $\SI{-116}{\dBm}$. A sub-GHz TWPA operating near the quantum limit could be an ideal solution for spin-qubit platforms, enabling goals such as frequency-multiplexed readout \cite{vesterinen2022sub}. KI-TWPAs are essential for reaching the full energy resolution of NIR/visible/UV single-photon kinetic inductance detectors operating at $\qtyrange{4}{8}{\giga\hertz}$ \cite{zobrist2019wide}, and sub-GHz TWPAs would bring the same advantage to far-IR single-photon detection with similar detectors \cite{day2024}.

Another application where improved sub-GHz amplifiers could yield large benefits is the search for QCD axion dark matter \cite{chaudhuri2024introducing, adams2022axion}. The QCD axion is a pseudo-Nambu-Goldstone boson, originally proposed as a solution to the strong CP-problem, that couples extremely weakly to the electromagnetic field tensor \cite{peccei1977cp,dine1981simple}. Over a wide mass range, $\sim \SI{1}{\pico\electronvolt}-\SI{1}{\electronvolt}$, the QCD axion has the characteristics required to be dark matter \cite{wilczek1978problem,weinberg1978new,sikivie2021invisible}. A common search method is the Sikivie cavity haloscope, which places a resonant cavity within a strong DC magnetic field to convert the axion of mass $m_a$ into a detectable microwave photon of frequency $\nu=m_ac^{2}/h$ through the inverse Primakoff effect \cite{sikivie1982axions,sikivie1983experimental}. Even with magnetic fields of $\SI{5}{\tesla}$ and several-hundred-liter cavities operating at a quality factor of $\sim 10^{5}$ , the expected signal power is on the order of $\SI{1e-22}{\watt}$, necessitating low-noise readout for detection \cite{ahn2024extensive}. 

Earlier work by the ADMX collaboration implemented resonant Microstrip SQUID Amplifiers (MSAs) over $\qtyrange{645}{680}{\mega\hertz}$ for Run 1A and resonant JPAs over $\qtyrange{680}{1020}{\mega\hertz}$ for Runs 1B-1C, enabling probes of QCD axions at or approaching the DFSZ sensitivity benchmark \cite{du2018search,braine2020extended,bartram2021search}. Development of a near-quantum-limited amplifier chain would allow DFSZ sensitivity at lower frequencies/masses \cite{chakrabarty2024low} and a TWPA could simplify the design and operation of axion searches \cite{ADMX:2021mio}.


In this paper, we present a KI-TWPA producing in excess of $\SI{30}{\dB}$ gain, and operating with $\numrange{2.5}{3.5}$ quanta of added noise while driven with $\qtyrange{300}{500}{\nano\watt}$ of pump power. Our results build on \citet{faramarzi2025near}, which introduced the device architecture and provided initial gain and noise characterizations in the $\qtyrange{500}{800}{\mega\hertz}$ frequency range. Our work extends the reach of these KI-TWPAs down to $\SI{300}{\mega\hertz}$ and presents additional characterizations under experimental conditions that are particularly relevant to axion searches. The gain-bandwidth, noise, and dynamic range of the KI-TWPA illustrate a powerful tool for precision measurements of sub-GHz signals.

Section \ref{sec:pxs-req} articulates the amplifier characteristics in the context of the Princeton Axion Search (PXS), and highlights the need for novel amplifier development. We present the gain performance of our device in Section \ref{sec:gain}, and present our noise measurement technique and results in Section \ref{sec:noise-results}. In preparation for integration with PXS, we present the dependence of noise on operating temperature and noise with an input circulator. In Section \ref{sec:error-analysis}, we analyze uncertainties in gain and noise measurements. In Section \ref{sec:conclusion} we conclude.
\section{First-Stage Amplifier Characteristics for the Princeton Axion Search}
\label{sec:pxs-req}


PXS, a Sikivie haloscope, will probe QCD axions over the $\qtyrange{200}{500}{\mega\hertz}$ range. First science will be accomplished with a $\SI{390}{\liter}$ cavity, immersed in a conduction-cooled solenoidal magnet generating a $\SI{5}{\tesla}$ magnetic field within the $\SI{740}{\milli\meter}$ bore. \citet{otto2026cavity} describe the cavity for the first operations, which will cover $\qtyrange{390}{480}{\mega\hertz}$. While the KI-TWPA described here is broadly applicable, its design is tailored to address performance targets for the first-stage amplifier connected to the cavity output. These performance targets inform characterizations performed in later sections of this paper. In addition to quantum-limited noise, the first-stage amplifier ideally has noise that gracefully degrades as the ambient operating temperature approaches $\SI{100}{\milli\kelvin}$; gain that is stable, reproducible, and sufficient to overcome the noise of follow-on amplifiers; bandwidth that covers the $\qtyrange{200}{500}{\mega\hertz}$ range with a minimal number of different bias conditions; and gain and noise performance that persist in $\mathcal{O}(\text{mT})$ magnetic fields.

Cavity haloscopes typically operate at temperatures of $\sim \qtyrange{50}{100}{\milli\kelvin}$ \cite{bartram2021search,ahn2024extensive}, substantially above the $\sim \SI{10}{\milli\kelvin}$ base temperature of a dilution refrigerator. This is partly due to the large thermal mass mounted to the mixing chamber. For example, the PXS haloscope requires cooling approximately $\SI{80}{\kilo\gram}$ on the mixing chamber of a Bluefors LD400 dilution refrigerator. Additional heat loads can arise from eddy currents in the cavity induced by mechanical vibrations and displacements in the high magnetic field. For a parametric amplifier, when these heat loads raise the mixing chamber temperature, the thermal occupation in the idler tone increases. (Section \ref{sec:device-description} discusses the relationship between signal and idler.) The idler thermal occupation is governed by Eq. \ref{eq:n_BE}:
\begin{equation}\label{eq:n_BE}
    n(\nu_{I})=\frac{1}{e^{h\nu_{I}/k_{B}T}-1}.
\end{equation}
Here, $\nu_{I}$ is the idler frequency and $T$ is the operating temperature. As detailed in \citet{malnou2024low}, the thermal occupation of the idler is a source of excess amplifier noise. At $\SI{100}{\milli\kelvin}$, the mean thermal occupation decreases from approximately $\num{10}$ photons at $\SI{200}{\mega\hertz}$ to $\num{0.6}$ photons at $\SI{2}{\giga\hertz}$. The frequency dependence motivates the use of a non-degenerate parametric amplifier with an idler frequency $\nu_I$ well above the $\qtyrange{200}{500}{\mega\hertz}$ signal frequency. This operation is unlike resonant JPAs deployed in higher-frequency QCD axion searches, where the signal and idler occupy the same resonant mode \cite{ahn2024extensive,bai2025dark,carosi2025search}. 
The KI-TWPA described here operates with idler frequencies from $\qtyrange{2.1}{2.6}{\giga\hertz}$ for our $\qtyrange{200}{500}{\mega\hertz}$ signal band. This separation between signal and idler frequencies reduces the thermal occupation of the idler mode and its contribution to system noise
\footnote{In addition to idler thermal noise, scaling the resonant JPA architecture used in higher-frequency haloscopes to PXS frequencies would come with the challenge of circulator loss. A circulator is needed to embed the intrinsically one-port JPA in the two-port microwave environment. The circulator loss, which tends to be larger at PXS frequencies relative to GHz counterparts, leads to additional noise. See Section \ref{sec:noise-measurement-results}.}.

In order for the first-stage amplifier to dominate the system noise, the gain must be sufficient to overcome the noise of follow-on amplifiers. At $\qtyrange{2}{5}{\giga\hertz}$, state-of-the-art semiconductor amplifiers operating at 4 Kelvin achieve input-referred noise of $\SI{1.5}{\kelvin}$ ($\sim \num{10}$ quanta) \cite{LNF1P2,LNF48}. Thus, $\sim \SI{15}{\deci\bel}$ gain is sufficient for quantum-limited amplifiers to overcome the noise of follow-on stages \footnote{To \textit{overcome} noise we mean the system noise can be well-approximated by the noise of the first amplifier. Consider, for example, a quantum-limited first-stage amplifier with $\SI{15}{\deci\bel}$ of gain and a second-stage amplifier operating at $\num{10}$ quanta, referred to its own input. Assume there is no loss between the two amplifiers. Using the Friis formula, it can be shown that the system noise referred to the first-stage input is only $\approx 1/3$ larger than the quantum noise.} \cite{chaudhuri2017broadband}. At the $\qtyrange{200}{500}{\mega\hertz}$ frequencies of PXS, state-of-the-art semiconductor amplifiers also achieve noise of $\SI{1.5}{\kelvin}$ ($\sim \num{100}$ quanta), so $\sim \SI{25}{\deci\bel}$ gain is needed \cite{LNF0P2,CITFL3}. 

A single cooldown of PXS is expected to last months, so it is important to have low gain drift and to account for any gain drifts in instrument calibration. Additionally, it is desirable to reproduce system parameters when re-scanning a potential axion candidate. The high-gain requirements of the sub-GHz regime and the adverse effects of gain drift on noise-temperature measurement accuracy were previously observed in the course of resonant MSA development for ADMX \cite{clarke2001microstrip,muck2001superconducting,kinion2011superconducting,o2019microstrip,khatiwada2021axion}. While the KI-TWPA described in this paper achieves the requisite gains in the majority of the PXS band, we build additional margin through simultaneous amplification and upconversion to the idler frequency \footnote{Simultaneous amplification and upconversion has been realized before, e.g., in Field Programmable Josephson Amplifiers \cite{lecocq2017nonreciprocal}.}. The gain drift and reproducibility are described in Section \ref{ssec:gain_drift}.

The data-taking phase of the axion search benefits from broad amplifier bandwidth \cite{braggio2022haloscope}. The narrowband cavity must be tuned to search a broad parameter space. Frequent tuning of a narrowband amplifier can introduce experimental complexity and increase downtime \cite{ahn2024extensive,kim2024parameter,bartram2021axion}. The KI-TWPA described here can cover the initial $\qtyrange{390}{480}{\mega\hertz}$ band with uniform noise performance in three tuning steps or fewer. See Section \ref{sec:noise-measurement-results} \footnote{Broader bandwidth-JPAs based on filter synthesis techniques have been realized recently\cite{kaufman2023josephson}. However, to our knowledge, they have not yet been applied to axion searches.}.

Even with field-cancellation coils, first-stage amplifiers in haloscopes can experience $\mathcal{O}(\text{mT})$ magnetic fields. The amplifier performance should gracefully degrade in these fields or the amplifier must have additional shielding \cite{ADMX:2011hrx,khatiwada2021axion}. Background fields can cause flux trapping and gain/noise degradation in Josephson-junction-based amplifiers, especially if the amplifiers possess SQUID elements. Cavity haloscopes using these amplifiers design sophisticated shields, consisting of field-cancellation coils as well as multiple layers of superconducting and ferromagnetic material, to protect their junction-based amplifiers \cite{brubaker2017first,ahn2024extensive}. Amplifiers based on the kinetic inductance nonlinearity in superconducting nitride thin films are more robust to background fields, which could simplify the required shielding. See refs. \cite{khalifa2023nonlinearity,xu2023magnetic,janssen2026temperature,boutan2026operation} for recent experimental measurements at several GHz and Section \ref{sec:noise-measurement-results} for further discussion.

Lastly, we note that as the dynamic range of the amplifier increases, so does immunity to electromagnetic interference that can saturate an amplifier and necessitate rebiasing \cite{kim2024parameter}. Additionally, increasing the amplifier's saturation power widens the usable range on a variable-noise source for noise calibration \cite{ahn2024extensive,guzzetti2025improved}. We present the output 1 dB compression point of our sub-GHz KI-TWPA in Section \ref{ssec:gain_meas}.

\section{Gain Characterization}
\label{sec:gain}

\subsection{Device Description}
\label{sec:device-description}
The architecture of the device characterized in this paper is described in detail in \citet{faramarzi2025near}. The device consists of an inverted microstrip transmission line \cite{shu2021nonlinearity}, with titanium nitride (TiN) center conductor, amorphous silicon (aSi) dielectric, and niobium skyplane. The TiN trace is $\SI{320}{\nano\meter}$ wide, $\SI{50}{\nano\meter}$ thick, and contains capacitive fingers to ground to produce a characteristic impedance near $\SI{50}{\ohm}$ \cite{chaudhuri2017broadband}. The microstrip length is $\SI{450}{\milli\meter}$ and meandered on a chip that is $\SI{2.5}{\milli\meter} \times \SI{25}{\milli\meter}$ \footnote{The device presented in this paper is from the same wafer as the device presented in \citet{faramarzi2025near}.}.

The TiN possesses a kinetic inductance that varies with applied current, and this provides the necessary nonlinearity for parametric gain. The amplifier described here operates in a four-wave-mixing mode \cite{agrawal2012nonlinear}. When a strong pump tone is applied alongside a weaker signal of a different frequency at the device input, power is transferred from pump to signal, creating amplification. Conservation of energy dictates that, in such four-wave mixing, a third tone must be created. This idler tone, of frequency $\nu_{I}$, is related to pump $\nu_{P}$ and signal $\nu_{S}$ frequencies by 

\begin{equation}
2\nu_{P}=\nu_{S}+\nu_{I}.
\end{equation}
To maximize gain, it is also necessary to conserve momentum, i.e. match phases among signal, pump, and idler tones \cite{chaudhuri2014simulation}. The phase matching is achieved by periodically modulating the capacitive fingers to add dispersion to the transmission line. The modulation allows the microstrip to approximate Eq. \ref{eq:phase-match} over a finite bandwidth, where the closer the real mismatch is to $\num{0}$, the greater the gain \cite{chaudhuri2014simulation,o2014resonant}.
\begin{equation}
    \label{eq:phase-match}
    \Delta k_{\mathrm{eff}}=\Delta k_{\mathrm{lin}}+2\Delta k_{\mathrm{NL}}\approx 0,
\end{equation}
where $\Delta k_{\mathrm{lin}}=k_S+k_I-2k_P$ is the linear wave-vector mismatch and $\Delta k_{\mathrm{NL}}$ represents the pump-induced nonlinear correction.

\subsection{Gain Measurement} 
\label{ssec:gain_meas}

Parametric gain in the KI-TWPA can be quantified by subtracting the signal transmission when no pump is applied from the transmission when applying a strong pump. This metric is referred to as the on/off gain \cite{howe2026kinetic,malnou2024low}. The on/off gain does not capture the contribution from loss mechanisms such as dissipation in the microstrip dielectric or the transition from the coaxial cable to the chip. Therefore, we present the true gain of the device as the difference in signal transmission between the pumped KI-TWPA and a lossless, through-line reference \cite{howe2026kinetic}.

Measurements are conducted in a Bluefors LD400 dilution refrigerator at $\SI{10}{\milli\kelvin}$. We conduct a dedicated cooldown where the TWPA is replaced with an SMA female-female barrel as a through-line reference. The system transmission (including cabling) is measured with a Keysight Technologies E5063A vector network analyzer.




The experimental setup for measuring TWPA response is shown in Appendix \ref{sec:gain-schematic}. The pump is provided by a Keysight Technologies N5183B analog generator and combined with the VNA probe (signal tone) using a Mini-Circuits ZFRSC-123-S+ power combiner. Starting with a pump power $\sim$20 dB below that needed for the onset of parametric gain, we ramp the pump power until gain is observed on the VNA. We refer to the resulting VNA signal transmission as the ``on'' transmission. The true gain is then determined by subtracting the through-line transmission from the ``on'' transmission,
\begin{equation}
  G_{\mathrm{true}}(\nu)=S_{21}^{\mathrm{on}}(\nu)  -S_{21}^{\mathrm{thru}}(\nu).
  \label{eq:true-gain}
\end{equation}  
All quantities in Eq. \ref{eq:true-gain} are expressed in decibels. 


\begin{figure*}
  \centering
  \includegraphics[width=\textwidth]{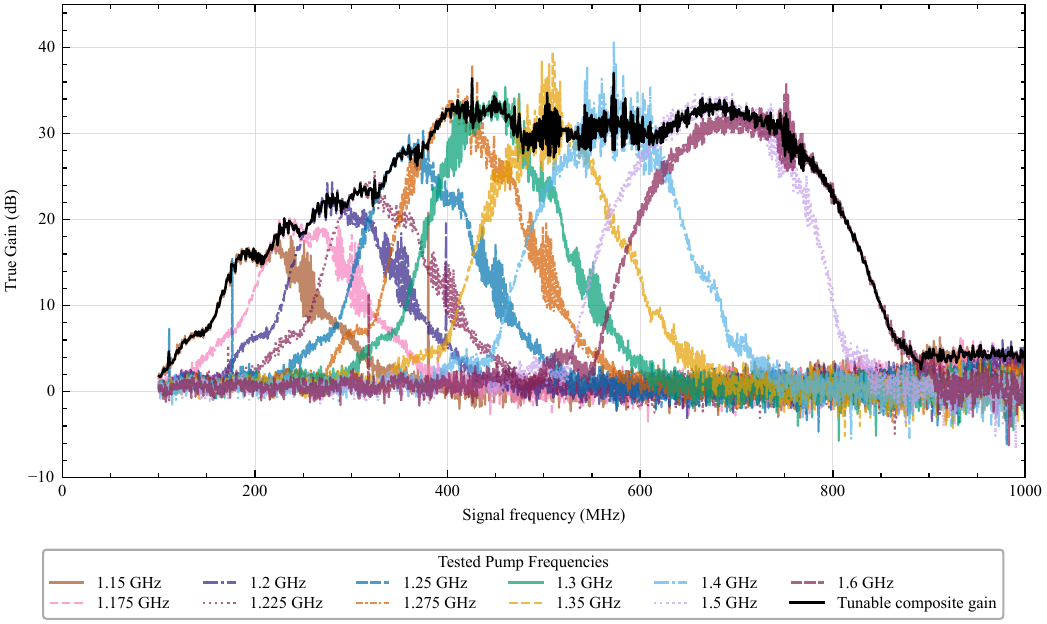}
  \caption{\label{fig:composite-gain}
    True gain of the TiN KI-TWPA. Each tested pump tone is plotted with its preferred operational condition. The pump powers at the device input range from $\qtyrange{300}{500}{\nano\watt}$. We represent the accessible gain for a single cold-hardware configuration with the \textit{Tunable composite gain} label. Each pump condition is plotted with the raw frequency resolution from the VNA, while the composite curve is binned into $\SI{1}{\mega\hertz}$ primarily to make the raw traces visible. The observed ripple can be reduced by operating with bias that produces less gain.}
\end{figure*}

Fig. \ref{fig:composite-gain} shows the true gain across different pump frequencies. The plotted curves represent the maximum pump power before the amplifier undergoes a superconducting-to-normal transition or the gain becomes ill-behaved. The maximum pump power ranges over $\qtyrange{300}{500}{\nano\watt}$ ($\SI{-35}{\dBm}$ to $\SI{-33}{\dBm}$), referred to the device input. The required pump power is a large improvement over the first KI-TWPAs produced a decade ago, for which the $\qtyrange{100}{200}{\micro\watt}$ pump powers could significantly heat external circuitry on the mixing chamber \cite{Eom2012,bockstiegel2014development}. The reduction in power is due to the reduced conductor cross sections as well as the use of TiN instead of NbTiN. During the course of testing the sub-GHz TWPA, the increase in mixing chamber temperature from applying the pump was $\leq \SI{1}{\milli\kelvin}$. 

To demonstrate that a single device presents broadband capability, we highlight the tunable-composite gain that can be selected from the relevant condition in the pump frequency space. The device shows appreciable gain over the $\qtyrange{200}{800}{\mega\hertz}$ band, with gain exceeding $\SI{30}{\deci\bel}$ at frequencies above $\SI{380}{\mega\hertz}$. While different pump frequencies are required to cover the $\qtyrange{200}{800}{\mega\hertz}$ span, the $\SI{3}{\deci\bel}$ bandwidth of a given pump frequency is greater than $\num{10}$\%. The fractional bandwidth is significantly larger than resonator-based parametric amplifiers currently deployed in QCD axion searches.

The ripple, observed at the gain peaks, generally arises from impedance mismatches with the external circuitry and imperfections in the microwave environment. The gain ripple possesses a frequency spacing of approximately $\SI{2}{\mega\hertz}$, corresponding to the electrical length of the entire microstrip. The phase velocity of the microstrip is $6 \times 10^{-3}c$. Outside of the $\SI{1.35}{\giga\hertz}$ and $\SI{1.4}{\giga\hertz}$ pump settings, the ripple in the tunable composite gain is largely confined to an amplitude of $\pm \SI{1.5}{\deci\bel}$. As described in \citet{klimovich2022traveling} (Eq. 4.12 and surrounding discussion), the gain ripple is highly nonlinear with pump power and can be reduced by operating at lower gains. (See also Eq. (B1) in \cite{howe2026kinetic}.) For prior TWPAs, operating at several GHz, the gain is lower and the ripple higher; see Table V of \cite{howe2026kinetic} for a summary. The high true gain is due, in part, to the low loss of the amorphous silicon dielectric \cite{defrance2024characterization,faramarzi2025near}. The microwave environment control that we achieve with our in-house coaxial cable assembly process further contributes to the high gain and low ripple. See Appendix \ref{sec:joe-ax-co} for details.

By sweeping the VNA probe power and adjusting the fixed attenuator prior to the power combiner, we measured the $\SI{1}{\deci\bel}$ compression point, referred to TWPA output, over a variety of pump conditions. The compression point was between $\SI{-56}{\dBm}$ and $\SI{-49}{\dBm}$, comparable to that reported in \citet{faramarzi2025near}.

\subsection{Gain Drift and Reproducibility}
\label{ssec:gain_drift}
We emphasize that the gain ripple is a stable feature of the TWPA. Neither the position in frequency nor the ripple amplitude measurably vary with time. As a concrete test of gain drift, we fixed the pump condition (frequency and power) and measured device transmission to acquire a reference trace. We maintained the bias and repeated gain measurements once per hour for at least four hours. Each hourly gain measurement was compared against the reference trace and the absolute values of the residuals were averaged over the $\SI{3}{\deci\bel}$ bandwidth. The test was repeated at several pump conditions that yield $\qtyrange{25}{30}{\deci\bel}$ gain. Over all conditions, the signal-frequency-averaged-residual was consistently $\SI{0.05}{\deci\bel}$ ($\num{1}$\%) or lower.

At the $\SI{1.5}{\giga\hertz}$ pump frequency with pump power set to yield $\SI{33}{\deci\bel}$ peak gain, we monitored the gain for consecutive days, yielding a mean-residual of $\SI{0.13}{\deci\bel}$. The gain was monitored across five days for the $\SI{1.3}{\giga\hertz}$ pump frequency with $\SI{30}{\deci\bel}$ peak gain, yielding a mean-residual of $\SI{0.15}{\deci\bel}$. A primary contribution to this residual could be the disconnect and reconnect of room-temperature test cables between days, which can cause a small shift in the pump transmission to the device. 

Although gain during a days-long cooldown was relatively constant, we observed larger shifts between cooldowns three months apart. The TWPA was biased with identical cabling and generator settings at two pump conditions ($\SI{1.35}{\giga\hertz}$, $\SI{+13}{\dBm}$ and $\SI{1.4}{\giga\hertz}$, $\SI{+13}{\dBm}$), each yielding peak gain of approximately $\SI{30}{\deci\bel}$. We compared the gain of each bias condition between the two cooldowns and the resulting mean-residual was $\SI{0.6}{\deci\bel}$ and $\SI{0.8}{\deci\bel}$, respectively.

One possible reason for the gain change, similar to the intra-cooldown monitoring, is a shift in the transmission of the pump between room temperature and device input. As the pump power increases, the gain increases and the center of the gain band shifts to higher frequency (cf. Fig. 3c in \citet{faramarzi2025near}). We observed such a change in the gain profiles for each bias condition between the two cooldowns. For phase-matched gains of $\SI{30}{\deci\bel}$, a $\SI{0.1}{\deci\bel}$ increase in pump power can result in a $\SI{0.8}{\deci\bel}$ increase in gain \cite{chaudhuri2014simulation}.  The thermal cycles on the coaxial cables and their interconnects and/or ambient temperature shifts changing the amplitude and phase response of the test cables could yield the difference in the pump transmission. 

The gain is reproducible across cryogenic systems. We have observed that, for a fixed pump condition on the analog generator, the gain of Fig. \ref{fig:composite-gain} changes by $\SI{0.3}{\decibel}$ or less when the mixing chamber temperature is raised to $\SI{300}{\milli\kelvin}$. We have tested another sub-GHz KI-TWPA from the same wafer (presented in Appendix \ref{sec:joe-ax-co}) in both the dilution refrigerator and a $^{3}$He/$^{4}$He sorption refrigerator operating at $\SI{270}{\milli\kelvin}$. In both systems, we achieve gains exceeding $\SI{30}{\decibel}$. When the device is biased to equivalent conditions (same pump frequency and producing comparable peak gain in the two systems), the difference in signal gain, averaged across the $\SI{3}{\deci\bel}$ bandwidth, is $\SI{2}{\deci\bel}$ or lower. The difference is comparable to the ripple amplitude. 
\section{Noise measurements}
\label{sec:noise-results}

\subsection{VTS Measurement Technique}
\label{sec:vts}

Several techniques are used to characterize near-quantum-limited amplifiers including shot-noise tunnel junctions \cite{spietz2003primary}, mesoscopic resistors \cite{bergeal2012mesoscopic}, and circuit quantum electrodynamics with a qubit \cite{macklin2015near}; see \citet{malnou2024low} for a comprehensive review. \citet{faramarzi2025near} used a Y-factor technique to switch between a hot load and cold load. For the noise measurements presented in this paper, we implement a variable temperature source (VTS) that allows noise calibration over an arbitrary number of input noise quanta set points \cite{simbierowicz2021characterizing,xu2023magnetic}.

The VTS consists of two copper stages, each $\SI{35}{\milli\meter}$ on a side and $\SI{6.4}{\milli\meter}$ thick. The top stage is thermally isolated from the bottom stage with four carbon fiber tubes of $\SI{5}{\milli\meter}$ OD, $\SI{3}{\milli\meter}$ ID, and $\SI{30}{\milli\meter}$ length. We create a weak thermal link with a  $\SI{70}{\milli\meter}$ length of $\num{30}$ AWG copper wire connecting the two stages. On the top stage, we install a $\SI{30}{\deci\bel}$ attenuator (Quantum Microwave, part number QMC-CRYOATT-30BLK), that is heat sunk with a copper L-bracket. A Lakeshore Cryotronics-calibrated ruthenium oxide thermometer (RX-102A-CD) is mounted to the L-bracket. The bottom of the VTS is mounted to the cold plate of the dilution refrigerator. 

To minimize conductive loading, the $\SI{30}{\deci\bel}$ attenuator is connected to the Still and MXC using $\SI{1.19}{\milli\meter}$-diameter and $\SI{2.19}{\milli\meter}$-diameter superconducting niobium-titanium (NbTi) coaxial cable, respectively. The use of an attenuator, as opposed to a $\SI{50}{\ohm}$ termination, allows one to inject signals into the TWPA from room temperature. This capability could be of particular use for monitoring system drifts in axion searches. In order to make thermal noise from higher temperature stages negligible, attenuators are installed at the $\SI{4}{\kelvin}$ and still stages preceding the VTS.

A $\SI{10}{\kilo\ohm}$ resistive heater is mounted to the top stage of the VTS. By applying current to the resistor, we can raise the temperature of the attenuator and increase the amount of noise output by the VTS. With $\SI{100}{\micro\watt}$ of applied power, we heat the VTS to just above 
$\SI{1.3}{\kelvin}$, while loading the $\SI{10}{\milli\kelvin}$ mixing chamber by less than $\SI{0.5}{\milli\kelvin}$. 

\subsection{Frequency-translation mode}
\label{sec:FT_mode}
In a typical noise measurement, the output from the VTS is amplified in several stages before acquisition by a spectrum analyzer. For linear amplifiers, the power in a frequency band of width $\Delta\nu$ centered at signal frequency $\nu_{S}$ is
\begin{equation}\label{eq:N_VTS_std}
P_{out}(\nu_{S})=G(\nu_{S}) \left( n(\nu_{S})+ \frac{1}{2} + N_{ex}(\nu_{S}) \right) h\nu_{S}\Delta\nu 
\end{equation}
where $G(\nu_{S})$ describes the total gain of the amplifier chain, $n(\nu_{S})$ is the thermal occupation (Eq. \ref{eq:n_BE}) at the VTS temperature, and $N_{ex}(\nu_{S})$ is the noise of the amplifier chain expressed in quanta. The factor of 1/2 represents zero-point fluctuations at the signal frequency. The equation is presented in photon-normalized units \cite{malnou2024low}. By sweeping the VTS temperature, we vary $n(\nu_{S})$. A linear fit of thermal occupation vs output power then yields $N_{ex}(\nu_{S})$, which has a minimum value of 1/2 per the SQL \footnote{The SQL is 1 photon of noise power per unit bandwidth, referred to the input of the first-stage amplifier. We use the convention that half of this power is attributed to zero-point fluctuations at the signal frequency and half to the amplifier itself.}.

Operating in a mode where the input and output are at the same frequency presents a particular challenge for quantum-limited readout in the PXS frequency range. Sub-GHz cryogenic semiconductor amplifiers have noise, expressed in quanta, that is significantly higher than equivalent amplifiers for GHz frequencies. As described in Section \ref{sec:pxs-req}, the first-stage amplifier must produce enough gain to overcome the semiconductor amplifier noise. As the first-stage noise approaches the quantum limit, the gain requirement becomes increasingly difficult to achieve. The KI-TWPA presented in this paper evades this constraint by using the frequency-translating (FT) mode introduced in \citet{faramarzi2025near}. 
The idler photons are in the $\qtyrange{2.1}{2.6}{\giga\hertz}$ frequency range, where commercial semiconductor amplifiers operating at $\sim \num{10}$ quanta are available. We thus use both amplification \emph{and} upconversion and read out the idler tone. The relationship between the output idler power and input signal noise quanta is
\begin{equation}\label{eq:N_VTS_FT}
P_{out}^{FT}(\nu_{I})=G^{FT}(\nu_{S}) \left( n(\nu_{S})+ \frac{1}{2} + N_{ex}^{FT}(\nu_{S}) \right) h\nu_{S}\Delta\nu.
\end{equation}
Here, $G^{FT}(\nu_{S})$ incorporates both the TWPA response (converting input signal photons into output idler photons) and the gain at idler frequencies of follow-on stages. By fitting the idler power versus the VTS noise quanta, we obtain the noise $N_{ex}^{FT}(\nu_{S})$ in frequency-translating mode. 

We emphasize reading out the idler is driven primarily by the need to circumvent the high noise of semiconductor amplifiers at PXS frequencies. If quantum-limited amplifiers can achieve sufficiently high gains with operational margin (for instance, through cascading), or sub-GHz semiconductor amplifier technology improves, the frequency-translating mode may become less advantageous. We also note that the ALPHA axion search, operating at tens of GHz, is considering an analogous frequency down-conversion approach based on kinetic inductance nonlinearity to interface the signal with standard junction-based measurement devices \cite{bai2026design}. 

\begin{figure*}[p]
  \centering
  \includegraphics[width=\textwidth,height=0.84\textheight,keepaspectratio]{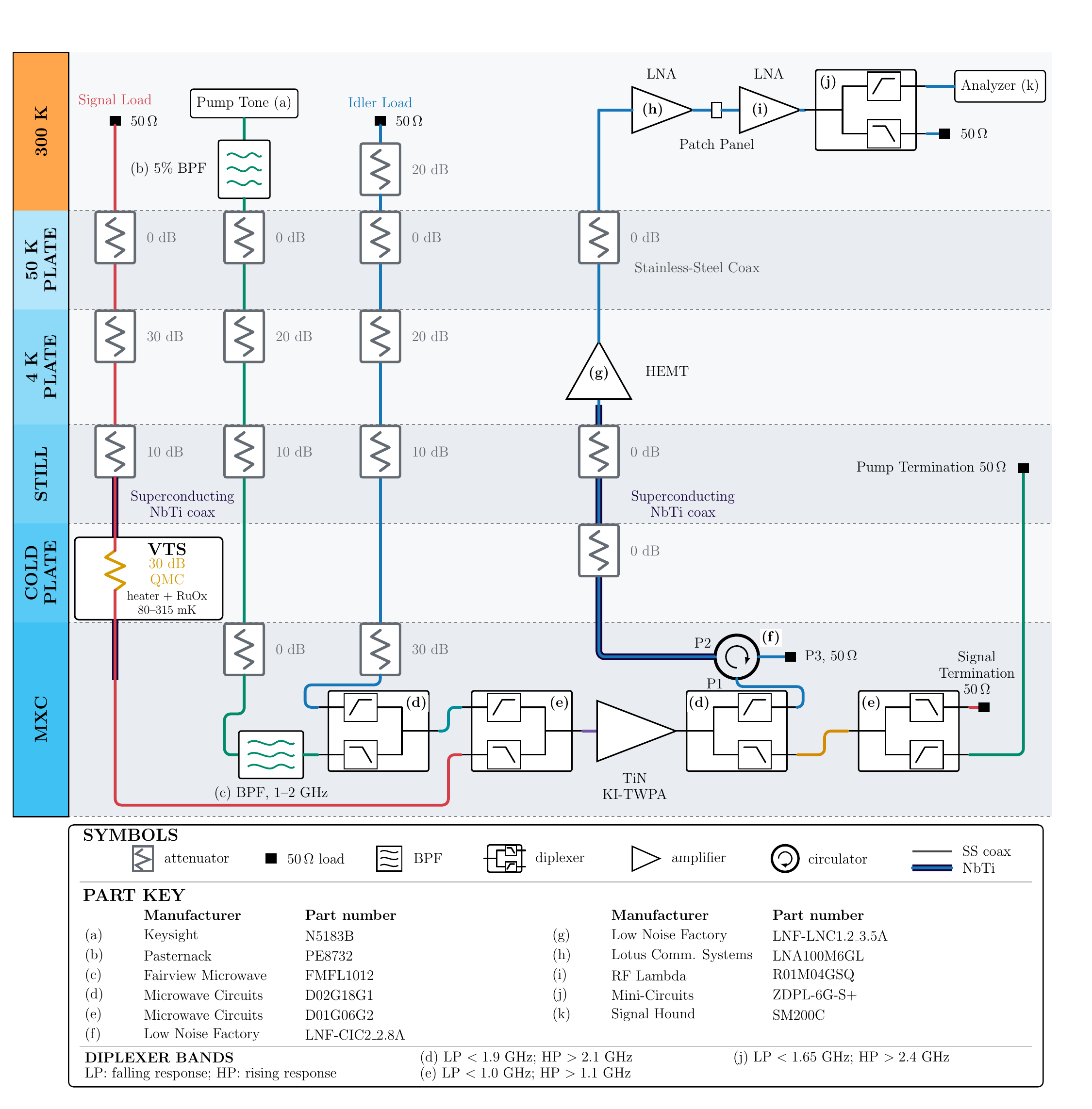}
  \caption{\label{fig:noise-setup}
    Experimental schematic for our variable temperature source (VTS)-based noise measurements in a Bluefors LD400 dilution refrigerator. This layout can be compared to that for ADMX \cite{khatiwada2021axion} and HAYSTAC \cite{jewell2023new}. The signal tone is labeled with the red path, the pump is labeled with the green path, and the idler is labeled with the blue path. The highlighted cables represent superconducting NbTi cable used for thermal separation between stages. The diplexers and bandpass filter are not optimized for cryogenic operation and were thus individually validated at $\SI{10}{\milli\kelvin}$. The connections between parts at the MXC are made with Mini-Circuits 0.086'' hand-formable cable.}
\end{figure*}

\subsection{Experimental Set-Up}
\label{sec:noise-set-up}

The noise measurement in frequency-translating mode is detailed in Fig. \ref{fig:noise-setup}. In this configuration, we combine the signal, pump, and idler tones with cold diplexers rather than warm power combiners. Cryogenic attenuators on the input channels mitigate the thermal noise from warmer temperature stages. At 4 Kelvin and below, we use Quantum Microwave attenuators (QMC-CRYOATT-XXNM) on the input pump line, because the internal quartz is an excellent thermal conductor to cool the center of the coax. This type of attenuator is also used at the MXC on the input idler line. All cold $\SI{50}{\ohm}$ terminations use the same construction (QMC-CRYOTERM-DC18NM).

During a noise measurement, a $\SI{50}{\ohm}$ termination is installed at room temperature on both the signal and idler line. The series of attenuators put the idler mode close to its vacuum state at the MXC, i.e. the thermal occupation is $n(\nu_{I}) \ll 1$. The idler line is also used to determine the gain band during amplifier biasing prior to the noise measurement. The pump tone is generated with a Keysight Technologies N5183B analog generator, and is conditioned with a Pasternack room temperature 5\% bandpass filter (PE8732) and Fairview Microwave cold bandpass filter (FMFL1012). The filters reject noise at the signal and idler frequencies that could contaminate the measurement.

We step from $\SI{0}{\volt}$ to our target in $\num{6}$ steps, and then return to $\SI{0}{\volt}$ by stopping at the $\num{5}$ intermediate points so that we can look for hysteresis/drift effects and long thermal time constants. In total, $\num{12}$ temperature set points are used (repeat $\SI{0}{\volt}$), with $\num{20}$ minute wait periods, resulting in a four hour measurement for every pump condition. Our tests did not find any discrepancy in measured noise between the heating and cooling directions for the VTS \footnote{We conducted measurements at 2-3 pump conditions per day. On occasion (roughly once every 1-2 days), the TWPA became unstable, resulting in a loss of amplification. The device was then re-biased and any affected data discarded. We did not observe a change in noise upon rebiasing.}.

The calibrated temperature sensor on the VTS is read out using the Bluefors Gen1 Temperature Controller, operating with $\SI{20}{\micro\volt}$ stimulus bound (Vmax excitation mode). To check that self-heating is not biasing the temperature measurement, we increased the stimulus to $\SI{200}{\micro\volt}$, observing an increase of less than $\SI{0.3}{\milli\kelvin}$. This increase is negligible, an order of magnitude below the thermometer calibration uncertainty \cite{BFG1TC,LakeShoreRX102A}.

The idler frequency output power spectrum is recorded using a Signal Hound SM200C analyzer. Our measurements use a resolution bandwidth of $\SI{5}{\kilo\hertz}$ with span adjusted depending on the pump condition. The result is a measurement of power $P_{out}^{FT}(\nu_{I})$ vs. input noise quanta $n(\nu_{S}) + 1/2$ for each frequency bin in the spectrum. A linear fit of this data, as per Eq. (\ref{eq:N_VTS_FT}), yields the noise $N_{ex}^{FT}(\nu_{S})$. The reference plane for all noise measurements is directly at the output of the VTS attenuator. Thus, losses between the VTS and TWPA input contribute to the measured amplifier noise $N_{ex}^{FT}(\nu_{S})$.

\subsection{Results}
\label{sec:noise-measurement-results}

We first present noise measurements at $\SI{10}{\milli\kelvin}$ operating temperature before discussing the effect of higher temperatures, magnetic shielding, and circulator prior to the TWPA.

\subsubsection{Baseline Measurements}

\begin{figure*}
  \centering
  \includegraphics[width=\textwidth]{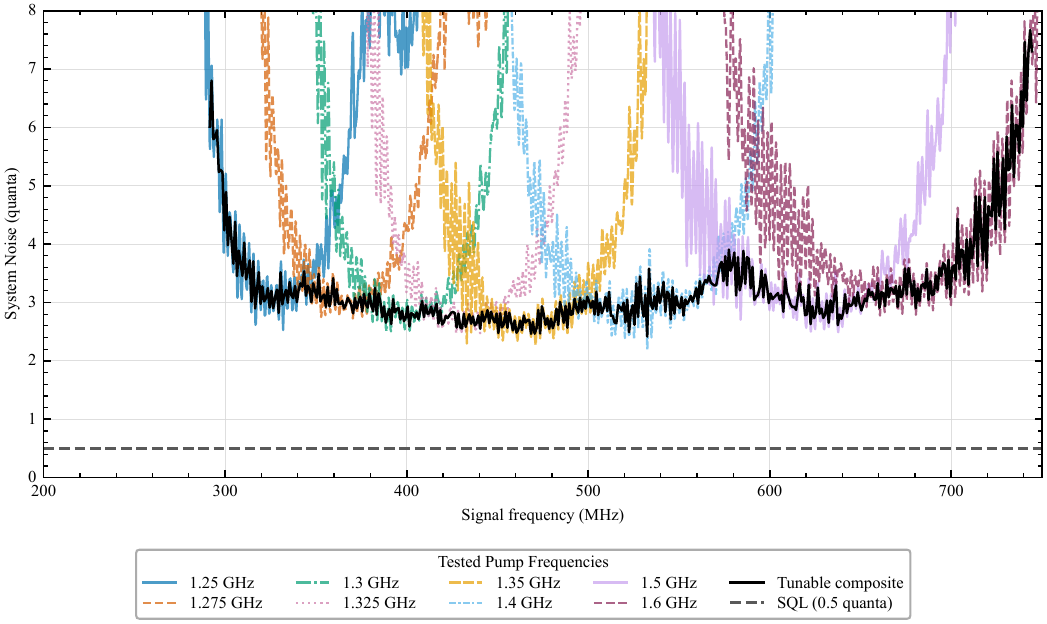}
  \caption{\label{fig:composite-noise}
    Our measured added system noise, $N_{ex}^{FT}(\nu)$, with the TiN KI-TWPA. The device demonstrates stable noise at 3 quanta referred to the output of the VTS. This noise level is achieved without magnetic shielding. For plotting clarity we split our data into $\SI{100}{\kilo\hertz}$ bins and present the mean value. The composite curve uses a $\SI{1}{\mega\hertz}$ bin width to demonstrate the accessible parameter space.}
\end{figure*}

For a range of pump frequencies, we set the pump power to yield $\qtyrange{20}{23}{\deci\bel}$ peak gain in the idler band. This gain is sufficient to overcome the noise of follow-on stages over bandwidths $\gtrsim$10\%. The VTS is swept over $\qtyrange{80}{315}{\milli\kelvin}$, and the noise measurement yields Fig. \ref{fig:composite-noise}. These measurements were conducted without ambient magnetic field shielding (TWPA operating in Earth's field), and were reproducible at the 5\% level, both within a cooldown and in two cooldowns spaced one month apart. For each frequency, our uncertainty is $\numrange{0.25}{0.33}$ quanta; see Section \ref{sec:noise-errors} for details. Similar to the gain results, we present the accessible noise regime in a single hardware configuration with the curve labeled \textit{tunable composite noise}. Over the $\qtyrange{300}{700}{\mega\hertz}$ band, we achieve noise between $\numrange{2.5}{3.5}$ quanta. While we achieve gain below $\SI{300}{\mega\hertz}$, the corresponding idler frequencies are sufficiently close to the $\SI{2.1}{\giga\hertz}$ band edge of the diplexer to cause performance degradation. The gap in uniform coverage near $\SI{580}{\mega\hertz}$ is due to observed hysteresis in the gain vs power behavior at $\SI{1.45}{\giga\hertz}$ pump in the noise-measurement setup. This anomaly was not present at other pump frequencies, so we elected not to take data there.

As with the gain, selecting appropriate pump conditions can optimize noise performance. For instance, to cover the $\qtyrange{390}{480}{\mega\hertz}$ range of the first PXS tuning rod, we can choose the $\SI{1.3}{\giga\hertz}$, $\SI{1.325}{\giga\hertz}$, and $\SI{1.35}{\giga\hertz}$ pump conditions in Fig. \ref{fig:composite-noise}. This would allow readout noise of 3 quanta or less with only three tuning steps. Fewer tuning steps may be possible by increasing the pump power.

\begin{figure}
  \centering
  \includegraphics[width=\columnwidth]{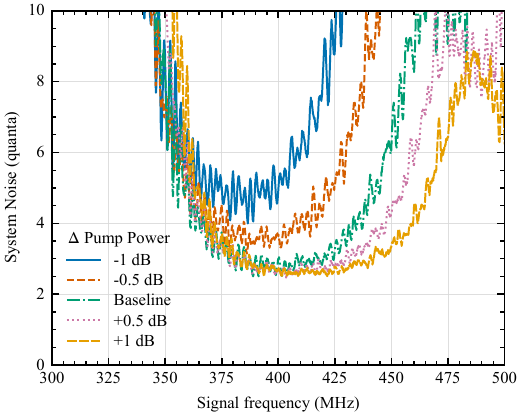}
  \caption{\label{fig:pump-power}
    Measured noise as a function of signal frequency for varied pump power configurations. The pump frequency is fixed at $\SI{1.3}{\giga\hertz}$. The ``Baseline'' trace corresponds approximately to a power of $\SI{300}{\nano\watt}$ at the device input.}
\end{figure}

In Fig. \ref{fig:pump-power}, we show the noise for various pump powers at $\SI{1.3}{\giga\hertz}$. The ``Baseline'' trace corresponds to the contribution to the tunable composite line in Fig. \ref{fig:composite-noise}. At low powers (e.g., blue curve, corresponding to 12 dB peak idler gain), the HEMT noise contributes significantly to the overall noise. As the pump power and gain increase (e.g., orange curve, corresponding to 30 dB peak idler gain), the bandwidth over which the amplified TWPA noise overwhelms the HEMT noise increases, and the minimum noise flattens out at 2.5 quanta. The noise above the quantum limit could be due to defects in the TWPA that create hotspots and thermal noise. It could also be due to poor thermalization in the internal components of the diplexers used to route the signal, pump, and idler tones. These diplexers are commercial components not optimized for cryogenic operation. 

\subsubsection{Dependence on Operating Temperature}
\begin{figure}
  \centering
  \includegraphics[width=\columnwidth]{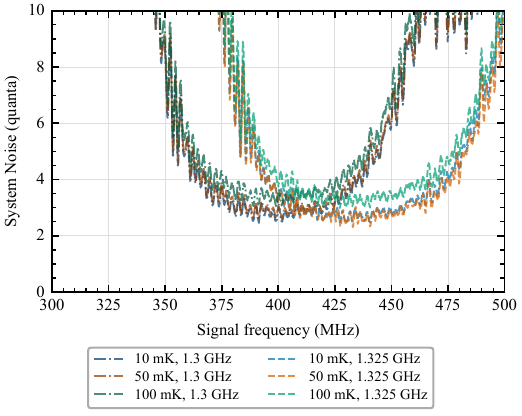}
    \caption{\label{fig:temperature} Increasing the temperature of MXC creates a higher thermal occupation in the idler band. The increased occupation from $\SI{10}{\milli\kelvin}$ to $\SI{50}{\milli\kelvin}$ is below our measurement resolution. Raising the temperature to $\SI{100}{\milli\kelvin}$ raises the measured system noise by $\num{0.6}$ quanta.}
\end{figure}

We stabilize the MXC at $\SI{50}{\milli\kelvin}$ and $\SI{100}{\milli\kelvin}$ and measure the amplifier noise. The pump frequency is set to $\SI{1.3}{\giga\hertz}$ and $\SI{1.325}{\giga\hertz}$, and we use pump powers identical to that in the tunable composite of Fig. \ref{fig:composite-noise}. While the occupation $n(\nu_{S}) + 1/2$ of a $\SI{400}{\mega\hertz}$ signal increases by a factor of $\num{8}$ between $\SI{10}{\milli\kelvin}$ and $\SI{100}{\milli\kelvin}$, the amplifier noise increases by a modest $\sim$30\%. The nominal increase between $\SI{10}{\milli\kelvin}$ and $\SI{50}{\milli\kelvin}$ is negligible and dominated by the measurement uncertainty. See Section \ref{sec:noise-errors}. The rise between $\SI{10}{\milli\kelvin}$ and $\SI{100}{\milli\kelvin}$ is, on average, 0.6 quanta, and close to that expected from the increased idler occupation: $n(\nu_{I})=2.6 \times 10^{-5}$ at $\SI{10}{\milli\kelvin}$, $n(\nu_{I})=0.14$ at $\SI{50}{\milli\kelvin}$, and $n(\nu_{I})=0.53$ at $\SI{100}{\milli\kelvin}$ for an idler frequency of $\SI{2.2}{\giga\hertz}$ \cite{malnou2024low}. The result is demonstrated in Fig. \ref{fig:temperature}.

\subsubsection{Sensitivity to Ambient Magnetic Field}
An amplifier's ability to operate in residual magnetic fields is important to axion searches due to the several-Tesla fields required to generate measurable axion signals. We re-test equivalent conditions with a two-layer $\mu$-metal shield around the dilution refrigerator's vacuum can to shield from Earth's magnetic field. We heat the device to $\SI{19}{\kelvin}$, above its superconducting transition temperature, install the shield, and cool back down. For test conditions, we use the $\SI{1.3}{\GHz}$ pump tone with the mixing chamber at $\SI{10}{\milli\kelvin}$ and $\SI{50}{\milli\kelvin}$, and the pump tone at $\SI{1.4}{\giga\hertz}$ while the mixing chamber is $\SI{10}{\milli\kelvin}$. We evaluate the residual (noise with shielding minus noise without shielding) in the sub-GHz gain band for each pump tone. We find the mean residual value to be $\num{0.02}$ quanta, and the standard deviation of the residual to be $\num{0.12}$ quanta. The influence of shielding is thus sub-dominant to the general measurement variance \footnote{\citet{howe2026kinetic} studies a NbTiN KI-TWPA operating at several GHz and similarly observes that the noise is unchanged when removing magnetic shielding.}. We have observed that the gain performance for a similar sub-GHz KI-TWPA (same materials and same microstrip trace width) degrades less than $\SI{0.1}{\deci\bel}$ when operating in a $\SI{1.5}{\milli\tesla}$ field perpendicular to the microstrip plane. In this test the field was applied after the device reached base temperature, and we did not cycle through $\text{T}_{c}$ while the field was applied.

\subsubsection{Input Circulator}

The resonator and the first-stage amplifier are typically operated at the same temperature in an axion experiment. Between them is an RF switch (to actuate between cavity and VTS for noise measurement), a circulator, and intermediate cabling. The circulator absorbs back-propagating noise in the amplifier chain; such noise could otherwise reflect off the cavity and appear at the amplifier input. The circulator is also used to calibrate the antenna coupling \cite{boutan2024axion}. At PXS frequencies, the circulator dominates the loss. Standard commercial circulators in this frequency range have an insertion loss of $\SI{-1}{\deci\bel}$, significantly worse than their counterparts at a few GHz \cite{QSTARCircs}.

To model the noise increase that PXS may have due to the circulator, we insert a circulator between the VTS and input diplexers. The circulator is not shown in Fig. \ref{fig:noise-setup}, but is placed between the VTS and component ``\textit{e}". The circulator we use is a specially-made unit from Quinstar that has insertion loss ranging between $\SI{-1.3}{\deci\bel}$ and $\SI{-2}{\deci\bel}$ and isolation ranging over $\qtyrange{15}{20}{\deci\bel}$ in the $\qtyrange{400}{450}{\mega\hertz}$ range. It is designed to operate in magnetic fields of up to $\SI{150}{\milli\tesla}$. The third port of the circulator is $\SI{50}{\ohm}$ terminated (part no. QMC-CRYOTERM-DC18NM), so that the circulator becomes an isolator. The total length of copper hand-formable cable between the VTS and first diplexer is approximately $\SI{600}{\milli\meter}$, similar to the cable length PXS will have between the cavity and first diplexer.

\begin{figure*}
  \centering
  \includegraphics[width=\textwidth]{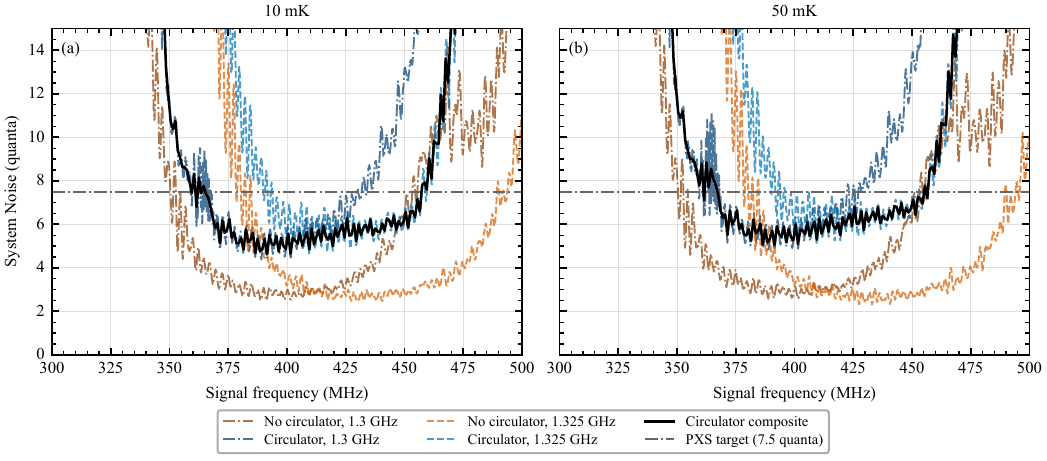}
  \caption{\label{fig:circulator}
    At both the base temperature of the dilution refrigerator, and the target operating point for PXS, an input circulator significantly increases the system noise. However, the result is within the PXS target for achieving DFSZ sensitivity. The feature near $\SI{365}{\mega\hertz}$, observed with the circulator and outside its vendor-specified band, is likely electromagnetic interference.}
\end{figure*}

We conduct noise measurements at $\SI{1.3}{\giga\hertz}$ and $\SI{1.325}{\giga\hertz}$ pump frequencies, at the same pump powers used in the tunable composite. We consider two temperatures, the $\SI{10}{\milli\kelvin}$ base temperature of the dilution refrigerator and the $\SI{50}{\milli\kelvin}$ target temperature of PXS. The result is shown in Fig. \ref{fig:circulator}. The presence of the circulator approximately doubles the noise, but the total noise, referred to VTS output, is still below the PXS target for DFSZ sensitivity (dash-dotted line)\footnote{The ADMX collaboration reports a similar amplifier noise target, $T_{n} \lesssim \SI{150}{\milli \kelvin}$, for reaching DFSZ sensitivity\cite{chakrabarty2024low}.}. 

A TWPA does not strictly need an input circulator in an axion search. If the  system-level impedance matching can be adequately controlled, so that reading out a reflective cavity does not render the TWPA unstable, and the intrinsic measurement backaction of the TWPA can be minimized and calibrated, then the circulator can be removed. In this case, the calibration of antenna coupling could be accomplished with a directional coupler. Our results showing increased noise with an input circulator also motivate efforts to realize sub-GHz TWPAs with intrinsic reverse isolation. Such TWPAs have already been implemented at several GHz \cite{ranadive2025travelling,malnou2025travelling,kow2026traveling}.

\section{Uncertainty analysis}
\label{sec:error-analysis}

\subsection{Uncertainty in Gain Measurement}
\label{sec:gain-errors}

A primary systematic error in our technique to measure true gain is establishing a consistent environment across the two cooldowns required to measure the through-line reference and the TWPA. Any shift in the environment transmission directly contributes to the evaluation of true gain in Eq. (\ref{eq:true-gain}). As a conservative estimate of the error, we use the shift in gain observed in the cooldowns three months apart in Section \ref{sec:gain} and round up to the nearest $\SI{1}{\deci\bel}$. We thus assign an uncertainty of $\SI{1}{\deci\bel}$ to our measurements of true gain.


\subsection{Uncertainty in the Noise Measurement}
\label{sec:noise-errors}

\begin{figure*}
    \centering
    \includegraphics[width=\textwidth]{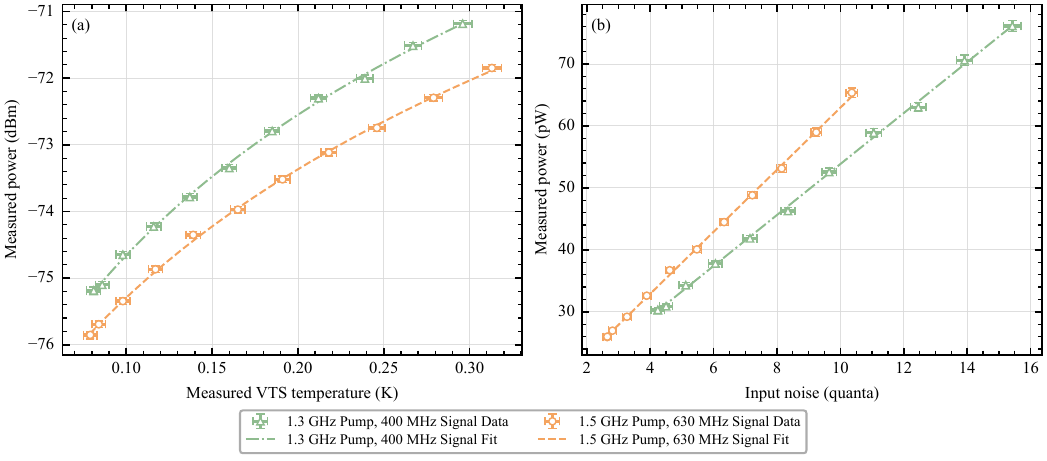}
    \caption{Example raw data from our measurement technique. Each data set is a single, fixed-frequency bin in the power spectrum and demonstrates the linear behavior between the input quanta and measured power. For our input quanta, we use the convention that includes the explicit 1/2 quanta contribution from vacuum fluctuations. The plotted samples are from the $\SI{1.3}{\giga\hertz}$ and $\SI{1.5}{\giga\hertz}$ pump tones. Panel A plots the measured VTS temperature and measured power at the signal analyzer, with the error bars from the direct measurement units. Panel B plots the power converted to Watts and converts the temperature to quanta with the propagated error bars. The linear fit in Panel B is from the orthogonal distance regression and what we use to extract the noise of the amplifier. We choose to transform that fit to the units for Panel A for this illustrative purpose.}
    \label{fig:linear-noise}
\end{figure*}

Our measurement, explained in Section \ref{sec:noise-set-up}, extracts the amplifier noise from the linear fit of the measured power against input noise quanta across our frequency spectrum. We configure the signal analyzer frequency spacing to have frequency-point spacing $<\SI{1}{\kilo\hertz}$ and a RBW of $\SI{5}{\kilo\hertz}$. An example is shown in Fig. \ref{fig:linear-noise}. Writing Eq. \ref{eq:N_VTS_FT} in slope-intercept form,

\begin{equation}
    \label{eq:N-ideal}
    N_{ex}^{FT}(\nu_{S})=\frac{b}{m}.
\end{equation}
Here, \textit{b} is the intercept in Watts, and \textit{m} is the slope in Watts per quanta. One advantage of this technique is the measurement accuracy does not depend on absolute power calibration, or absolute gain or loss from downstream components. The accuracy is limited primarily by the ability to measure temperature of the input load, and the linearity of the amplifier chain including the spectrum analyzer \cite{o2019microstrip}. The typical tradeoff is this simple approach requires slow measurements. As discussed in Section \ref{sec:noise-set-up}, each measurement is approximately $\num{4}$ hours, but the gain stability discussed in \ref{ssec:gain_drift} is sufficient to conduct these measurements with linearity, seen in Fig. \ref{fig:linear-noise}, without re-calibrating gain during the measurement.

To determine the error in the extracted noise, we evaluate the uncertainty in input noise quanta and uncertainty in measured power, determine the best fit with an orthogonal distance regression, and use the covariance matrix to extract our propagated uncertainty.

The input noise quanta is a function of frequency and temperature. We can treat the frequency resolution of the Signal Hound SM200C as exact because, even without a GPS lock, the internal oscillator operates with drift $\pm$ $\SI{2e-8}{\text{}\per \text{day}}$, which makes the propagated error in the frequency sub-dominant to temperature uncertainty by several orders of magnitude. The constraint on temperature is the dominant uncertainty among uncertainties derived from datasheet specifications. The input quanta are from both the population from the physical temperature of the VTS, as well as residual population from warmer stages since the cables run from room temperature to amplifier input. As shown in Fig. \ref{fig:noise-setup}, we use a total of $\SI{70}{\deci\bel}$ of fixed attenuation to reduce the number of photons from preceding temperature stages. Without accounting for additional attenuation from cable loss, the residual additional quanta is subdominant to our VTS blackbody input quanta by two orders of magnitude. Ignoring the residual can overestimate the system noise: however, the effect is subdominant to our thermometry uncertainty such that ignoring the discrepancy does not change our findings.

The calibrated ruthenium oxide (RuOx) temperature sensor on the VTS has an uncertainty of $\pm$ $\SI{4}{\milli\kelvin}$ \cite{LakeShoreRX102A}, and the Bluefors temperature controller has a fractional uncertainty of 1\% in measured resistance for the current range associated with our $\SI{20}{\micro\volt}$ stimulus \cite{BFG1TC}. We treat the RuOx and temperature controller uncertainties as uncorrelated and add them in quadrature. We propagate the resulting uncertainty in temperature to an uncertainty in input noise quanta:
\begin{equation}
    \label{eq:temp-uncert}
    \sigma_{n}(\nu) = \frac{\frac{h\nu}{k_{B}{T}}e^{h\nu/k_{B}T}}
        {\left(e^{h\nu/k_{B}T}-1\right)^{2}}
        \frac{\Delta T}{T}
\end{equation}

The linearity of the spectrum analyzer is difficult to constrain because it depends on the specific high speed ADC as well as the analog and digital electronics used for signal conditioning. Even if the response is a function of frequency, the analysis routine only requires a linear response at each fixed frequency. Because our configuration uses a fixed span, resolution bandwidth, and reference level, we can provide a conservative amplitude uncertainty by attributing the variance of the gain drift measurements in \ref{ssec:gain_drift} to the spectrum analyzer ($\SI{0.05}{\deci\bel}$). We avoid further noise contributions by operating more than $\SI{30}{\deci\bel}$ above the analyzer noise floor and more than $\SI{60}{\deci\bel}$ below the $IP_{3}$ compression point. 


Applying the fit covariance matrix to the error propagation of Eq. \ref{eq:N-ideal} results in Eq. \ref{eq:N-prop-error}.

\begin{equation}
    \label{eq:N-prop-error}
    \sigma_{N_{ex}^{FT}} = \sqrt{\left(\frac{b}{m^{2}}\right)^{2}\sigma_{m}^{2}
        + \left(\frac{1}{m^2}\right)\sigma_{b}^{2} - \frac{2b}{m^3}\sigma_{bm}}
\end{equation}

The propagated uncertainty in the extracted system noise is $\numrange{0.25}{0.33}$ quanta derived from the specified uncertainties in VTS thermometry and our estimated uncertainty in measured output power. Basic quality-of-fit metrics demonstrate that our data follows the linear model (Eq. \ref{eq:N_VTS_FT}) well. We have $R^2 \geq 0.99$ for $>99\%$ of the measurements. For these points, we have reduced-$\chi^2 \leq 10^{-3}$ so the propagated effects do not appropriately express our uncertainty. These metrics do not constrain systematic effects that preserve linearity. One example of such a systematic effect is loading that makes the input quanta differ from the assumption that our VTS is an ideal blackbody at its measured temperature.

\section{Conclusion}
\label{sec:conclusion}

We have measured a TiN KI-TWPA producing gain between $\qtyrange{200}{800}{\mega\hertz}$, with peak gain in excess of $\SI{30}{\deci\bel}$ above $\SI{380}{\mega\hertz}$ and output 1 dB compression point between $\SI{-56}{\dBm}$ and $\SI{-49}{\dBm}$. Further, we have demonstrated noise characterization for this device operating in frequency-translating mode. Our measurements determined the system noise with this device is $\numrange{2.5}{3.5}$ quanta for the $\qtyrange{300}{700}{\mega\hertz}$ range. Further testing is required to determine whether the noise is primarily intrinsic to the KI-TWPA or caused by losses between the VTS and device input. We demonstrated that the noise level is unaffected by Earth's magnetic field within our cryostat and increases modestly at mixing chamber temperatures up to $\SI{100}{\milli\kelvin}$. We characterized the noise with an input circulator. While the circulator doubles the input-referred noise, the setup meets the scientific requirements for the Princeton Axion Search to reach DFSZ sensitivity. Future directions include a comprehensive study of gain/noise dependence on magnetic fields up to the critical field of niobium, including $1/f$ stability \cite{dantsker1997high,stan2004critical}; replacing the normal metal diplexers with lithographed superconducting filters to reduce loss and associated noise; constructing a sub-GHz KI-TWPA with reverse isolation; and realizing squeezed states, as pioneered by HAYSTAC, to enhance axion search sensitivity \cite{backes2021quantum,zheng2016accelerating}. 


\begin{acknowledgments}
    The authors thank Farzad Faramarzi and Sasha Sypkens for the initial screening of the device described in this paper. We thank Pat Bradshaw and Steve Lowe for their critical assistance in the Princeton Physics student machine shop. We thank Kent Irwin and Lindley Winslow for useful discussions. This work has been supported by Simons Foundation grant number MP-TMPS-00002996 and the ``Table-top experiments for fundamental physics'' program, sponsored by the Gordon and Betty Moore Foundation, Simons Foundation, Alfred P. Sloan Foundation, and John Templeton Foundation. We gratefully acknowledge support from the Princeton Innovation Fund for New Ideas in the Natural Sciences. This material was based upon work supported by the U.S. Department of Energy, Office of Science, Office of High Energy Physics, under Award Number DE-SC0025942. Research and fabrication development was carried out at the Jet Propulsion Laboratory under a contract with the National Aeronautics and Space Administration (80NM0018F0610). 
\end{acknowledgments}

\appendix
\section{Experiment Set-up Schematics}
\label{sec:schematics}

\subsection{Gain Measurement}
\label{sec:gain-schematic}
\begin{figure}
  \centering
  \includegraphics[width=\columnwidth]{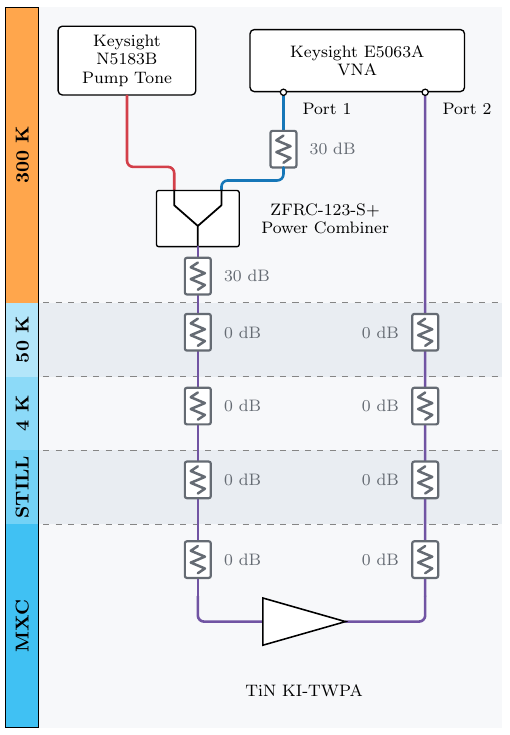}
  \caption{\label{fig:gain-setup}
    Experimental set up for gain screening in our Bluefors LD400 Dilution Refrigerator. The stimulus port from the VNA creates the reference for both the signal and idler tones, and the analog generator creates the pump. All tones are combined and enter the cryostat along a single cold cable. Connections at the MXC are made with 0.086" Mini-Circuits hand-formable cable.}
\end{figure}

Gain measurements, discussed in Section \ref{sec:gain} and demonstrated in Figure \ref{fig:composite-gain}, are made according to the schematic in Figure \ref{fig:gain-setup}. We use a Keysight Technologies E5063A Vector Network Analyzer as the signal source and a Keysight Technologies N5183B to generate the pump tone. The VNA probe tone and synthesizer pump tone are combined at room temperature using a Mini-Circuits ZFRC-123-S+ power combiner to enter the cryostat via a common line. The device is mounted on the mixing-chamber plate, which operates between 10 and 11 mK throughout the testing process. We use stainless-steel coax (UT-085-SS-SS from Amphenol/CarlisleIT) between temperature stages and Mini-Circuits 086-6SM+ hand-formable coax at the mixing chamber. The input and output coax lines are thermalized using fixed attenuators (XMA 2082-6418-00-CRYO) at each stage. The $\SI{0}{\deci\bel}$ attenuators simultaneously provide thermalization to the coax center conductor, and preserve enough signal so that we are able to conduct the measurement without room temperature amplifiers. 

A typical VNA trace for gain measurement consists of 10,001 points across 100 MHz-4 GHz, $\SI{70}{\kilo\hertz}$ IF bandwidth, and 500 averages. The acquisition time for such a trace is slightly under 280 seconds.

\subsection{Noise Measurement}
\label{sec:noise-schematic}

Noise measurements discussed in Section \ref{sec:noise-measurement-results} and presented in Fig. \ref{fig:composite-noise}, \ref{fig:pump-power}, and \ref{fig:temperature} are conducted with the apparatus shown in Fig. \ref{fig:noise-setup}. Fig. \ref{fig:circulator} includes a circulator, not shown in Fig. \ref{fig:noise-setup}, in between the VTS and the diplexer labeled component ``\textit{e}".

\section{In-house Coaxial Cable Fabrication}
\label{sec:joe-ax-co}

The quality of the external microwave environment can affect TWPA performance. Impedance mismatches can result in gain ripple and limit gain through the early onset of parametric oscillation \cite{klimovich2022traveling}. To improve quality control of our impedance environment while reducing costs, we developed the capability to fabricate semi-rigid coaxial cable assemblies in-house. These assemblies demonstrated measurable improvements in our KI-TWPA performance. In this appendix, we describe the fabrication process. 

\begin{figure}
    \centering
    \includegraphics[width=\columnwidth]{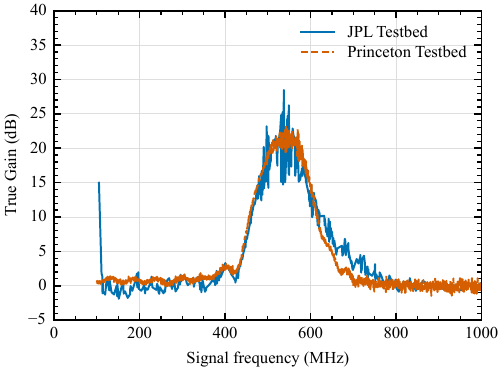}
    \caption{\label{fig:testbed-gain}True gain measured for the device in this manuscript with $\SI{1.4}{\giga\hertz}$ pump tone. The pump power from the Princeton testbed is selected to match the JPL testbed gain, which is less power than we used in Fig.~\ref{fig:composite-gain}. Both curves are presented with no smoothing.}
\end{figure}

Unless the microwave environment is well-controlled, two different testbeds can yield different TWPA performance. An example is shown in Fig. \ref{fig:testbed-gain}. The amplifier discussed in the main text was screened initially at JPL (blue) to confirm basic functionality. We observed that when the amplifier was biased to equivalent conditions in our system, the amplifier demonstrated less ripple (orange). The amplifier could also be operated at higher gains by further increasing pump power.

\begin{figure*}
    \centering
    \includegraphics[width=\textwidth]{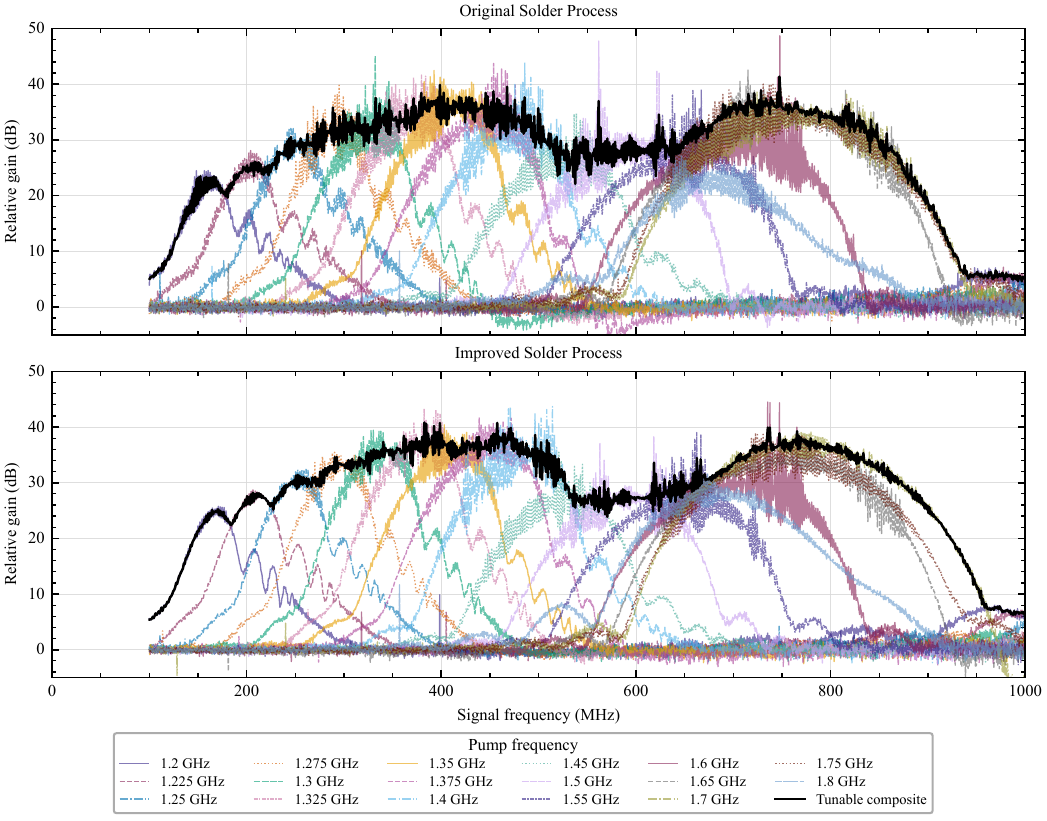}
    \caption{Another KI-TWPA, not otherwise discussed in this paper, demonstrated cleaner transmission after we replaced some of our stainless steel cabling. The new cabling featured an improved solder technique and surface preparation. We plot the signal transmission at maximum pump power, relative to the signal transmission with pump $\sim$20 dB below the onset of amplification. The sensitivity of the KI-TWPA to the impedance environment revealed the improvement. Both figures are presented with raw VNA resolution.}
    \label{fig:R4C45-gain}
\end{figure*}

Another demonstration of the amplifier's sensitivity to the impedance environment was evident when we were screening a separate device. See Fig. \ref{fig:R4C45-gain}. The original screening used an earlier in-house fabrication process, which began to show degradation (corrosion and cracking) in the solder joints after $\sim$25 cooldowns. 
We replaced the cables with our updated fabrication process and again saw improvement in the gain. The only change was the stainless steel cabling.


We fabricate $0.086"$ stainless-steel, NbTi, and copper coax in addition to $0.047"$ NbTi and copper coax. We are able to connect with either SMA or $\SI{2.92}{\milli\meter}$ connectors. Fabricating $0.047"$ cable is possible, but more challenging due to the material fragility. We fabricate the cables on a Hardinge lathe in the Student Shop at Princeton University. The lathe is equipped with a 0.1-thousandths digital readout (DRO), and we use a ThinBit carbide cutter that is 30-thousandths wide to trim the cables. All cables were bent using a 3D printed tool designed in-house, which gives us a convenient mechanism to create arbitrary bends without exposing the cable to any materials that could introduce crimps or kinks. Material-specific fabrication details are included in the relevant sections.

\subsection{Stainless Steel}
\begin{figure}
    \centering
    \includegraphics[width=0.85\columnwidth, angle=180]{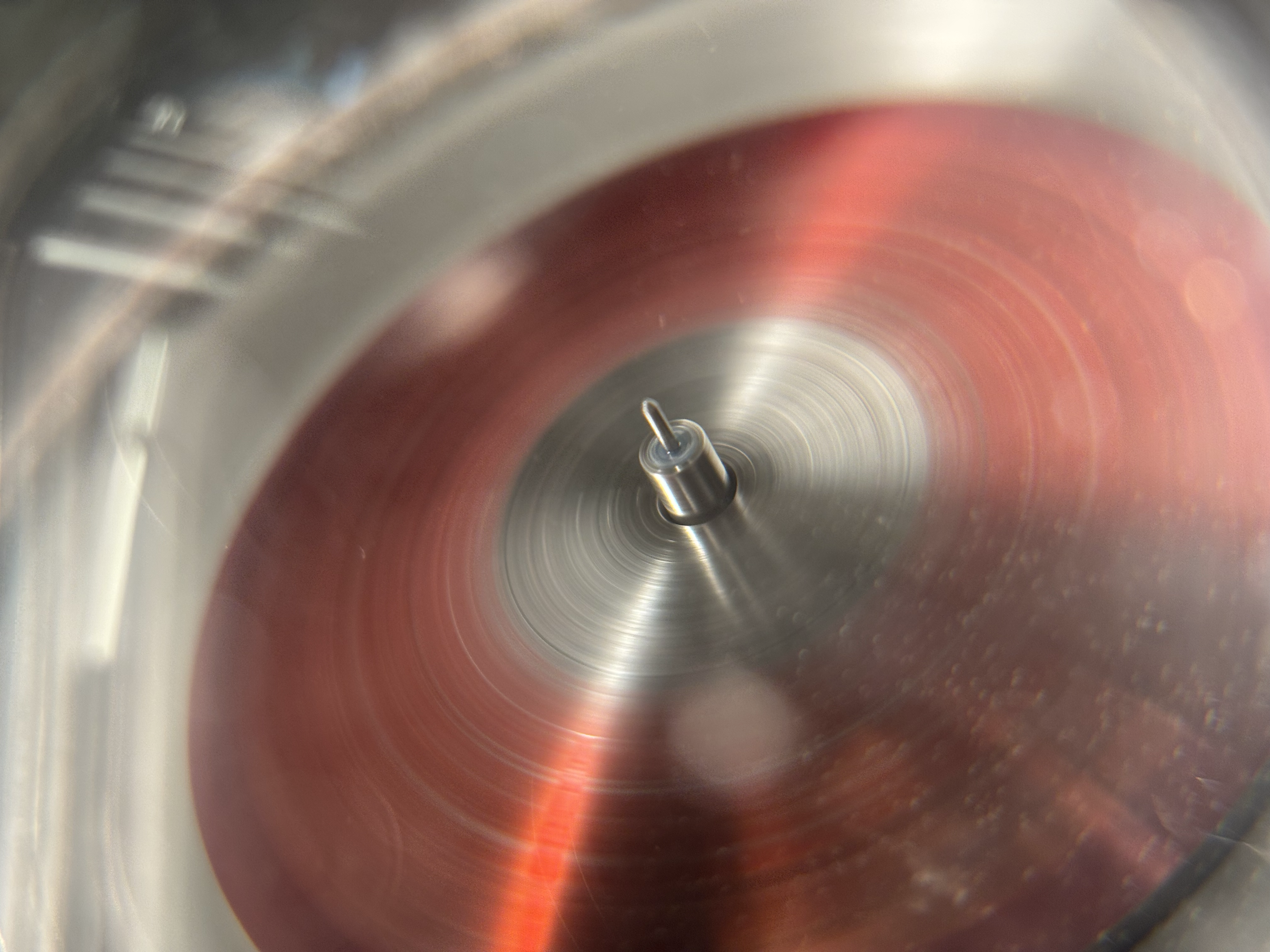}
    \caption{\label{fig:coax-lathe} An example stainless steel cable trimmed on the student shop lathe.}
\end{figure}

We source our stainless steel stock from Amphenol/CarlisleIT, and use Radiall 9401-1583-010 soldered connectors. The bulk stock is cut using an angle grinder cutter in  a stand, with gentle clamping force on the coax. The angle grinder ensures a sharp cut without deforming the cable out of round, or bending the center pin. The cut cables are spun on the lathe around 3000 RPM, and trimmed to the connector specification. The DRO is zeroed each end of the cable by dusting the end of the cable, which becomes our dominant fabrication tolerance. We can set a conservative estimate of $\SI{15}{\micro\meter}$ ($\num{0.5}$-thousandths of an inch) as our trim tolerance, weighted toward cutting beyond the target length. We trim the cable with no cutting lubricants, and take particular care to set our cutter to be at the equator of the cable. We position the cutter with a slight cutting bias toward the leading edge to ensure the cut edge of the cable remains as clean as possible. Once trimmed, we remove the stainless steel grounding jacket with nippler pliers, and then carefully cut the dielectric by hand using a razor blade while the lathe spins. Cutting without mechanical advantage helps ensure the center conductor remains unperturbed. To finish the side, a small hand-file is used to dust off oxide, and bring the center conductor to a point. The polishing effect of the file improved the solder flow, so we typically expose a small additional amount of the cable to polish before cutting the next cable. An example result is shown in Fig. \ref{fig:coax-lathe}.

Once machined, we sonicate the cables in isopropyl alcohol with a small amount of heat for 20 minutes total, alternating the submerged side every few minutes. We then prepare the connectors on a clean benchtop, and insert the cable into the compression fittings. We use a lint-free cotton swab to apply a zinc-free stainless steel solder flux to the cable and solder with a high silver content solder. When done in batches, the process can be efficient. PXS was able to have one researcher fabricate the 48 cables for the experiment instrumentation in a total of about 20 working hours.

\subsection{NbTi}
We tested many NbTi recipes and found for our most sensitive applications, crimped connectors could not achieve our target return loss performance. Our two preferred schemes are $0.086"$ NbTi from EZ-Form cable with either SMA or $\SI{2.92}{\milli\meter}$ connectors from EZ-Form, or $0.047"$ NbTi with Coaxicom $\SI{2.92}{\milli\meter}$ connectors. The bulk cutting and trimming steps are largely the same as stainless steel, except more care needs to be taken to adjust the tightness on the collet to ensure that the cable is neither deformed out of round, nor imprinted with shallow grooves from the contact points from the collet. This is particularly delicate for the smaller diameter cable. For $0.047"$ cables, Hardinge sells a special diameter collet that fits these cables, and the lathe should spin as fast as possible. NbTi oxidizes very quickly, so immediately placing the cables in an isopropyl alcohol bath after polishing is a good practice to reduce the oxide growth.

\begin{figure}
    \centering
    \includegraphics[width=0.70\columnwidth]{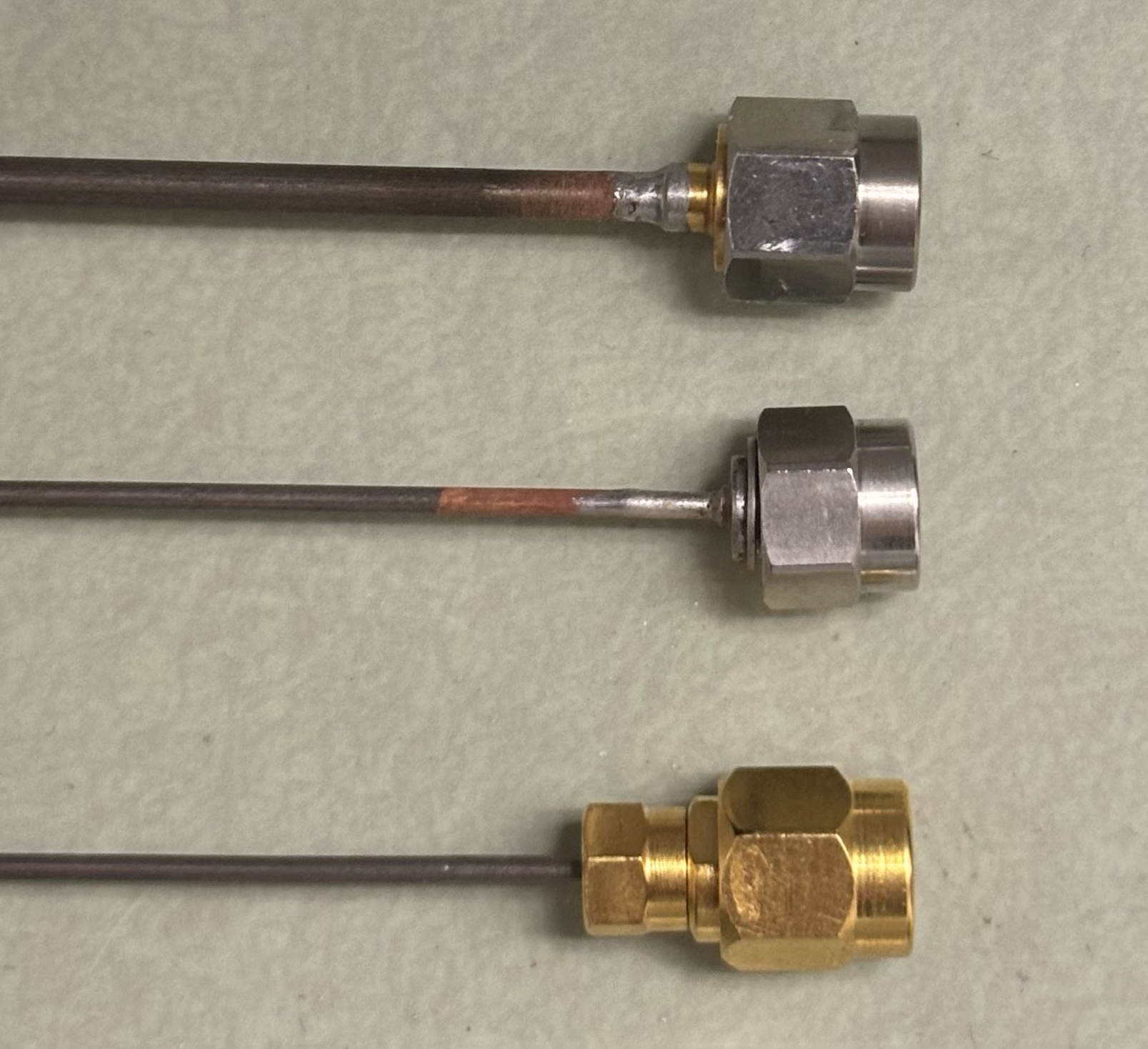}
    \caption{\label{fig:nbti-trip} Three example NbTi cables from our testing, on top a 0.086" cable with Radiall SMA connector, center 0.047" with Coaxicom SMA connector, and bottom 0.047" with Intelliconnect SMA connector.}
\end{figure}

NbTi will not adhere to solder without surface prep. We follow the copper-plating procedure from \cite{debry2018note}. For practical implementation we found it effective to use a current-limiting resistor to give us the very small current draw required to plate the $0.047"$ cable sufficiently slowly. Part of our preference for the Coaxicom $\SI{2.92}{\milli\meter}$ connectors for $0.047"$ cable is the connector geometry does not require plating the center conductor, and the compression gets tighter as the connector cools down. Examples of $0.086"$ and $0.047"$ cable are shown in Fig. \ref{fig:nbti-trip}.

\subsection{Copper}
Copper cable fabrication is similar to NbTi, without the need to plate the cable. Machining the cable is delicate because the material is soft. We typically reserve machining copper cables for long runs that require semi-rigid cables, because many of our testing applications show no meaningful degradation from hand-formable copper cables.\\

\subsection{On the Use of Fixed Attenuators}
One potential method for mitigating impedance mismatches from semi-rigid cable is to use cold fixed attenuators. However, especially for the pump tone, the resulting dissipation can heat the cryogenic environment. For example, adding a $\SI{10}{\deci\bel}$ fixed attenuator on the pump input line at the MXC in Fig. \ref{fig:noise-setup} results in up to $\SI{5}{\mu\watt}$ of dissipation. Localized heating in the attenuator can be a source of excess noise. For multichannel systems using many TWPAs, the cumulative thermal load may raise the MXC temperature and result in increased idler thermal noise (see Sections \ref{sec:pxs-req} and \ref{sec:noise-measurement-results}).

Fixed attenuators have nonzero return loss and can also contribute to impedance mismatches and gain ripple. We conducted a test in which, on the TWPA input line, the $\SI{0}{\deci\bel}$ attenuators at $\SI{4}{\kelvin}$ and still were replaced with $\SI{20}{\deci\bel}$ and $\SI{10}{\deci\bel}$ attenuators, respectively (part nos. 2082-6418-XX-CRYO from XMA). Coax with the improved soldering process was used, and the $\SI{30}{\deci\bel}$ attenuator after the power combiner was removed (see Appendix \ref{sec:gain-schematic}). Across PXS frequencies, the gain ripple with cold attenuators was consistently higher than that shown in the bottom panel of Fig. \ref{fig:R4C45-gain}; at some pump frequencies, the maximum achieved gain was a few dB lower. This observation motivates careful selection of commercial cryogenic microwave components in experiments using TWPAs, as well as the comprehensive production of cryogenic return-loss data for such components.

Additionally, fixed attenuators increase the amount of pump power that must be sourced from room temperature. One must be careful not to exceed the source capabilities of low-noise synthesizers (often limited to $+\qtyrange{15}{20}{\dBm}$ unless the instrument is specially-configured). As the frequency increases, the cable loss increases. Efforts to realize quantum-limited K-band TWPAs \cite{shu2021nonlinearity} may then benefit from well-engineered semi-rigid cable with $\SI{2.92}{\milli\meter}$ connectors.

\bibliography{references}

\end{document}